\documentclass[10pt,journal]{IEEEtran}

\usepackage{latexsym}
\usepackage{graphicx}
\usepackage{epsfig}
\usepackage{subfigure}
\usepackage{array}
\usepackage{amsmath}
\usepackage{amssymb}
\usepackage{color, soul}
\usepackage{amsthm}
\usepackage{enumerate}
\usepackage{algpseudocode}
\usepackage{algorithm}

\newcommand{\bg}{\textbf g}

\newcommand{\bW}{\textbf W}

\newcommand{\bG}{\textbf G}
\newcommand{\bV}{\textbf V}

\newcommand{\bw}{\textbf w}

\newcommand{\ba}{\textbf a}
\newcommand{\bb}{\textbf b}
\newcommand{\bv}{\textbf v}

\newcommand{\bp} {\begin{proof}}
\newcommand{\ep} {\end{proof}}

\theoremstyle{plain} 
\newtheorem{thm}{Theorem}

\newtheorem{prop}{Proposition}
\theoremstyle{definition}

\def\bal#1\eal{\begin{align}#1\end{align}}

\begin{document}

\title{Comments on "On Favorable Propagation in Massive MIMO Systems and Different Antenna Configurations" \cite{Wu-17}}

\author{S. Loyka, M. Khojastehnia

\vspace*{-1\baselineskip}

\thanks{S. Loyka and M. Khojastehnia are with the School of Electrical Engineering and Computer Science, University of Ottawa, Canada, e-mail: sergey.loyka@uottawa.ca}
}

\maketitle

\begin{abstract}
It is shown that the condition of Theorem 1 in \cite{Wu-17} never holds in practice and that Theorem 2 is incorrect under the stated condition. Extra assumptions or/and modifications are needed to make the conclusions of Theorem 1 and 2 above valid, which are provided below.

\end{abstract}


\section{Introduction}

In this note, we show that the key condition of Theorem 1 in \cite{Wu-17} never holds in practice and that Theorem 2 in \cite{Wu-17} is incorrect under the stated condition. This is done by identifying key gaps in their proofs and via a counter-example. Following this, we show that the conclusions of Theorem 1 and 2 do hold if additional assumptions and modifications are introduced. While these additional assumptions are restrictive, they are justified by the physics of radio wave propagation and do hold in many practical scenarios.

Unless stated otherwise, we use the same assumptions and notations as in \cite{Wu-17}, and assume that all expectation terms exist and bounded; all expectations are with respect to the distribution of all random variables inside of the expectation operator.

\section{Theorem 1 in \cite{Wu-17}}

For completeness, we state below Theorem 1 in \cite{Wu-17}, where $M$ and $L$ are the number of BS antennas and multipath components, $\bg_i$ is the channel vector of (single-antenna) user $i$, $\alpha_{ri}$ is the propagation gain of multipath component $r$ for user $i$, which is also $ri$-th entry of matrix $\bV$  (see \cite{Wu-17} for the channel model and definitions of other related symbols).

\begin{thm}[\cite{Wu-17}]
Let $\bW \in \mathbb{C}^{M\times L}$ be the steering matrix that consists of $L$ column vectors
\bal
\bw_l  = \ba_{BS}(AoA_l)
\eal
where $\bw_l \in \mathbb{C}^{M\times 1}$ is the $l$-th column vector of $\bW$, and $\ba_{BS}(AoA_l)$ represents the steering vector of the BS antennas for the $l$-th path. Then the condition
\bal
\label{eq.1.15}
\frac{1}{M} \sum_{m=1}^M \mathbb{E}\{ w_{mr}^* w_{ms}\} \to 0
\eal
is a sufficient condition to guarantee that
\bal
\label{eq.1.16}
\mathbb{E} \left\{\frac{1}{M}  \bg_i^H \bg_k \right\} \to 0
\eal
and then the FP condition holds\footnote{Note that \eqref{eq.1.16} is different from the standard definition of favorable propagation, where there is no expectation: $\frac{1}{M}  \bg_i^H \bg_k \to 0$ , see e.g. \cite[(2)]{Wu-17}\cite[Ch.7]{Marzetta-16}-\cite{Ngo-14}. The former, i.e. $\mathbb{E} \left\{\frac{1}{M}  \bg_i^H \bg_k \right\} \to 0$, does not imply the latter, i.e. $\frac{1}{M}  \bg_i^H \bg_k \to 0$, without certain additional assumptions. For example, the random variable $\frac{1}{M}  \bg_i^H \bg_k$ may take values $\pm 1$ with equal probability so that its mean is zero, but the RV itself does not converge to zero in any sense.}, provided only that $\alpha_{ri}^*\alpha_{sk}$ and $w_{mr}^* w_{ms}$ are uncorrelated, that is provided only that steering matrix $\bW$ and wireless channel gain matrix $\bV$ are uncorrelated.
\end{thm}

To see the gap in the formulation of this Theorem, which invalidates its conclusion, observe that
\bal
\frac{1}{M} \sum_{m=1}^M \mathbb{E}\{|w_{mr}|^2\} = \frac{1}{M} \mathbb{E}\{|\bw_{r}|^2\} = 1
\eal
where $\bw_r = [w_{1r},..,w_{Mr}]^T$, $|\bw_{r}| = (\sum_m |w_{mr}|^2)^{1/2}$ is the Euclidean norm of $\bw_{r}$, and the last equality is due to the normalization
\bal
\label{eq.norm}
|\bw_r|^2 = \sum_{m=1}^M |w_{mr}|^2  = M
\eal
as in e.g. (23), (28), and (35) in \cite{Wu-17}. Hence, the condition \eqref{eq.1.15} (which is \cite[(15)]{Wu-17}) is never satisfied for $r=s$ under this normalization, since
\bal
\frac{1}{M} \sum_{m=1}^M \mathbb{E}\{w_{mr}^*w_{mr}\}=  \frac{1}{M} \mathbb{E}\{|\bw_{r}|^2\} = 1
\eal
for any $M$, does not matter how large (note that the $r=s$ terms are present in  $\mathbb{E}\{z\} = \mathbb{E} \left\{\frac{1}{M}  \bg_i^H \bg_k \right\}$, see  \cite[(12), (13)]{Wu-17}, and hence are essential in the proof of Theorem 1 in \cite{Wu-17}). Thus, the conclusion of Theorem 1 never holds under this normalization. Furthermore, \cite[(26), (31), (38)]{Wu-17} also fail for $r=s$, invalidating the respective conclusions.

It should be noted that the normalization in \eqref{eq.norm} above, as well as in \cite[(23), (28), and (35)]{Wu-17}, is not arbitrary but quite essential to make the definition of orthogonality, as in e.g. \eqref{eq.1.16} or \cite[(16), (19)]{Wu-17},  consistent. Indeed, let $\ba, \bb$ be 2 real-valued vectors. The angle $\theta$ between these vectors can be found from
\bal
\cos \theta = \frac{\bb^T\ba}{|\bb||\ba|} = \frac{1}{M} \bb^T\ba
\eal
where the last equality holds under proper normalization, $|\ba|=|\bb| =\sqrt{M}$. Since orthogonality means $\cos \theta = 0$, this implies the following definition of orthogonality
\bal
\frac{1}{M} \bb^T\ba  = 0
\eal
and explains the appearance of $\frac{1}{M}$ in this definition, which is essential when $M\to \infty$.

To fix this gap of Theorem 1, an extra assumption and a modification are needed, as indicated below.

\begin{prop}
Consider the setting of Theorem 1 in \cite{Wu-17} and, in addition, assume that the propagation channel vectors of different users are orthogonal to each other "on average", i.e.
\bal
\label{eq.p1.1}
\mathbb{E}\{\bv_i^H\bv_k\} = \sum_r \mathbb{E}\{\alpha_{ri}^*\alpha_{rk}\} = 0,\ i\neq k.
\eal
Then, the condition
\bal
\label{eq.p1.2}
\frac{1}{M} \sum_{m=1}^M \mathbb{E}\{w_{mr}^*w_{ms}\} \to 0,\ \forall\ r \neq s,\ \mbox{as}\ M \to \infty
\eal
implies asymptotic orthogonality "on average"\footnote{we stress again that \eqref{eq.p1.3} does not imply $\frac{1}{M}\bg_i^H \bg_k \to 0$, i.e. the FP according to the standard definition, without some additional assumptions.}.
\bal
\label{eq.p1.3}
\mathbb{E}\{z\} = \frac{1}{M} \mathbb{E}\{\bg_i^H \bg_k \} \to 0,\ i \neq k,\ \mbox{as}\ M \to \infty.
\eal
\begin{proof}
Observe the following
\bal
\label{eq.p1.p1}
\mathbb{E}\{z\} &= \frac{1}{M} \sum_{r=1}^L \sum_{s=1}^L \mathbb{E}\{\alpha_{ri}^*\alpha_{sk}\} \sum_{m=1}^M \mathbb{E}\{w_{mr}^*w_{ms}\} \\
\label{eq.p1.p2}
&= \frac{1}{M} \sum_{r\neq s} \mathbb{E}\{\alpha_{ri}^*\alpha_{sk}\} \sum_{m=1}^M \mathbb{E}\{w_{mr}^*w_{ms}\} + \sum_{r} \mathbb{E}\{\alpha_{ri}^*\alpha_{rk}\}\\
\label{eq.p1.p3}
&= \frac{1}{M} \sum_{r\neq s} \mathbb{E}\{\alpha_{ri}^*\alpha_{sk}\} \sum_{m=1}^M \mathbb{E}\{w_{mr}^*w_{ms}\} \\
\label{eq.p1.p4}
&\to 0
\eal
where the summations are over all possible values of $r$ and $s$, $r\neq s$ means that the $r=s$ terms are excluded; \eqref{eq.p1.p1} is \cite[(13)]{Wu-17} (due to $\alpha_{ri}^*\alpha_{sk}$ and $w_{mr}^*w_{ms}$ being uncorrelated); \eqref{eq.p1.p2} follows from the normalization in \eqref{eq.norm}, \eqref{eq.p1.p3} follows from \eqref{eq.p1.1},  and \eqref{eq.p1.p4} follows from \eqref{eq.p1.2}.
\end{proof}
\end{prop}

Compared to \cite[Theorem 1]{Wu-17}, the orthogonality condition in \eqref{eq.p1.2} is imposed for $r\neq s$ only, and the extra condition in \eqref{eq.p1.1} is added, which can be motivated by independent scattering environment around each user, see Fig. 1 below, which induces independent small-scale fading around each user and hence $\alpha_{ri}$ and $\alpha_{rk}$ are independent, $i\neq k$. This is a standard assumption in the popular Kronecker correlation model, where geographically-separated antennas experience independent scattering and fading \cite{Kermoal-02}\cite{Heath-19}\cite{Costa-10} as well as in the keyhole channel model \cite{Marzetta-16}\cite{Heath-19}\cite{Costa-10}-\cite{Levin-08}. Finally, we point out that \eqref{eq.p1.3} holds without expectation (i.e. the standard definition of the FP, where the convergence is in probability) if \eqref{eq.p1.1} and \eqref{eq.p1.2} hold without expectation as well.

\section{Theorem 2 in \cite{Wu-17}}

For completeness, we state below Theorem 2 in \cite{Wu-17}.

\begin{thm}[\cite{Wu-17}]
Let the complex channel gain matrix, $\bG$, be expressed as the matrix product
\bal
\bG= \bW\bV
\eal
where $\bW$ and $\bV$ are the steering matrix and the wireless channel angle of arrival signal gain matrix, respectively where the element $\alpha_{lk}$ of $\bV$ characterizes the amplitude and phase changes of the signal on the $l$-th path for the $k$-th user, including path loss, time delay and Doppler frequency shift. Then, if the $|\alpha_{lk}|$, $l = 1,...,L$ and $k=1,...,K$ are bounded, that is if the condition $|\alpha_{lk}| < \infty$ , for all $l$ and $k$ is satisfied, the condition,
\bal
\label{eq.1.18}
\frac{1}{M} \sum_{m=1}^M \mathbb{E}\{ w_{mr}^* w_{ms}\} \to 0
\eal
is sufficient to guarantee that the FP condition is satisfied\footnote{see Footnote 1.}, that is
\bal
\label{eq.1.19}
\mathbb{E} \left\{\frac{1}{M}  \bg_i^H \bg_k \right\} \to 0
\eal
holds.
\end{thm}

This Theorem is "proved" in \cite{Wu-17} via the following key inequality:
\bal
\label{eq.1.21}
\mathbb{E}\{z\} \le \frac{C_{\alpha}^2}{M} \sum_{r=1}^L \sum_{s=1}^L \sum_{m=1}^M \mathbb{E}\{ w_{mr}^* w_{ms}\}
\eal
which is \cite[(21)]{Wu-17}, where
\bal
\mathbb{E}\{z\} = \mathbb{E} \left\{\frac{1}{M}  \bg_i^H \bg_k \right\} = \frac{1}{M} \sum_{r=1}^L \sum_{s=1}^L \sum_{m=1}^M \mathbb{E}\{ w_{mr}^* w_{ms} \alpha_{ri}^*\alpha_{sk}\}
\eal
which is \cite[(12)]{Wu-17}, and
\bal
\label{eq.1.20}
|\alpha_{ri}| \le |\alpha|_{max} = C_{\alpha} < \infty
\eal
for all $r, i$, which is \cite[(20)]{Wu-17}.

To demonstrate that this Theorem is incorrect, we show that \eqref{eq.1.21} above (i.e. \cite[(21)]{Wu-17}) and its usage in the proof of Theorem 2 are incorrect in several different ways.

1. Observe that both sides of \eqref{eq.1.21} are complex numbers in general. However, complex numbers cannot be ordered \cite{Bronshtein-86} (i.e. one cannot say that one complex number is larger or smaller than the other, in the same sense as for real numbers), unless their difference is a real number, which is not the case here in general. Hence, this inequality cannot be correct in the general case.

2. Even if all terms of \eqref{eq.1.21} are real (or, more generally, the difference of two sides is real), \eqref{eq.1.18} does not imply \eqref{eq.1.19} via \eqref{eq.1.21}, since $\mathbb{E}\{z\}$ can be negative (as the scalar product of 2 vectors), so that $\mathbb{E}\{z\}=-1$ is possible in \eqref{eq.1.21} even if its right-hand-side is zero, e.g. if  \eqref{eq.1.18} holds.

3. It can be further demonstrated, via a counter-example, that \eqref{eq.1.21} is not correct in general under the stated assumption of boundedness in \eqref{eq.1.20}, even if all terms are real. To this end, consider the following example satisfying the conditions of Theorem 2: set $i=1$, $k=2$, and
\bal
w_{mr}=\alpha_{r1}=\alpha_{r2}=a_r,\ r=1...L,\ m=1...M
\eal
with any $L \ge 2, M \ge 1$. Further assume that random variables $\{a_r\}$ are independent of each other (pair-wise independence is sufficient), zero-mean and have bounded support, e.g. set $a_r= \pm 1$ with equal probability, so that $C_{\alpha}=1$. Straightforward computation gives:
\bal
\mathbb{E}\{z\} = \sum_{r,s} \mathbb{E}\{a_r^2 a_s^2\} = L^2
\eal
yet the right-hand-side of \eqref{eq.1.21} is
\bal
C_{\alpha}^2 \sum_{r,s} \mathbb{E}\{a_r a_s\} = \sum_{r}  \mathbb{E}\{a_r^2 \}= L < L^2 = \mathbb{E}\{z\}
\eal
clearly violating the inequality \eqref{eq.1.21}, even for all real variables.

4. Finally note, from Section II, that the right-hand side of \eqref{eq.1.21} does not necessarily converge to zero: if $r=s$, the respective summation terms add up to $L C_{\alpha}^2 >0$ for any $M$. Also note that \eqref{eq.1.18} is the same as \eqref{eq.1.15}, which never holds for $r=s$, as it was explained above.

In view of the above discussion, a question arises whether the asymptotic orthogonality "on average" can be ensured when $\bW$ and $\bV$ are not independent (correlated with) of each other. The following Proposition answers this question in affirmative.

\begin{prop}
\label{prop.orth.2}
Consider the channel as in \cite[(17)]{Wu-17}, i.e. $\bG=\bW\bV$, and assume that $\alpha_{ri}=a_r b_i$ for all $r, i$, where random variables $b_i$ are zero-mean, independent of each other and of $a_r$ and $w_{mr}$, for all $m, r, i$. Then, the orthogonality "on average" holds:
\bal
\label{eq.orth.M}
\mathbb{E}\{z\} = \frac{1}{M} \mathbb{E}\{\bg_i^H \bg_k \} = 0,\ i \neq k.
\eal
\begin{proof}
Under the stated assumptions, the following holds:
\bal\notag
\mathbb{E}\{z\} &= \frac{1}{M} \sum_{r=1}^L \sum_{s=1}^L \sum_{m=1}^M \mathbb{E}\{\alpha_{ri}^*\alpha_{sk} w_{mr}^*w_{ms}\} \\
\label{eq.p2.p2}
&= \frac{1}{M} \sum_{r=1}^L \sum_{s=1}^L \sum_{m=1}^M \mathbb{E}\{a_r^* a_s w_{mr}^*w_{ms}\} \mathbb{E}\{b_i^* b_k\} = 0
\eal
where the last 2 equalities are due to the independence of $b_i$ from the other  variables and from each other, $\mathbb{E}\{b_i^* b_k\} = \mathbb{E}\{b_i^*\} \mathbb{E}\{b_k\} =0$.
\end{proof}
\end{prop}

The setting of Proposition \ref{prop.orth.2} can be slightly broadened by assuming that $b_i$ are not correlated with each other and with $a_r^* a_s w_{mr}^*w_{ms}$, instead of being independent.

Note that the orthogonality in \eqref{eq.orth.M} holds for any $M$, not just asymptotically, and that $\alpha_{ri}$ and $w_{mr}$ are not necessarily independent (uncorrelated) of each other, since $a_r$ and $w_{mr}$ are not. This is due to the joint dependence of $a_r$ and $w_{mr}$ on the angle of arrival of incoming multipath component $r$. Note also that no assumption of bounded support is needed here, just the existence and boundedness of respective expectation terms.

The assumed factorization of propagation coefficients $\alpha_{ri}=a_r b_i$ is justified by the physics of radio wave propagation and is reminiscent of the Kronecker MIMO channel correlation model, where the impacts of scattering environments abound the transmitter (user) and receiver (base station) are separated and independent of each other, which was confirmed experimentally in certain environments \cite{Kermoal-02}, and which is a popular model for MIMO channels  \cite{Heath-19}\cite{Costa-10}. This factorization is also observed for rank-deficient or keyhole channels, see e.g. \cite[Ch. 7]{Marzetta-16}\cite[Ch. 3]{Heath-19}\cite{Levin-08}; a detailed analysis of
keyhole channels can be found in \cite{Chizhik-02}, including the factorization property. In our case, $b_i$ model scattering around each user, which are independent of each other due to separate locations, and $a_r$ model scattering around the base station, which is independent of that of the users, due to the same reason. This is illustrated in Fig. 1.

\begin{figure}[t]
    \begin{center}
        \includegraphics[width=3in]{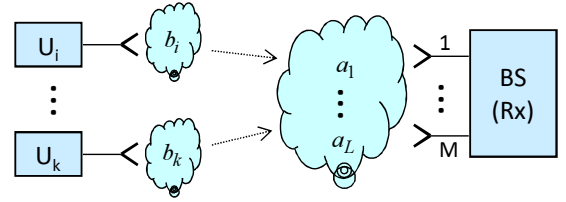}
    \end{center}
    \caption{Propagation environment with different and independent scatterings around each user and base station. It is the BS scattering that induces $L$ multipath components arriving to the BS antenna array.}
\end{figure}


\end{document}